\documentclass[prl,twocolumn,showkeys,superscriptaddress,nofootinbib]{revtex4}
\usepackage{times}
\usepackage{amsmath}
\usepackage{graphicx}
\usepackage{epsfig}
\usepackage{dcolumn}
\usepackage{bm}
\usepackage{ulem}
\usepackage{color}
\usepackage{array}
\usepackage{multirow}
\usepackage[utf8]{inputenc}
\usepackage{CJK}

\newcommand{\BR}{{\cal B}}

\newcommand{\jpsi}{J\kern-0.15em/\kern-0.15em\psi\kern0.15em}

\newcommand{\psip}{\psi(2S)}

\newcommand{\EE}{e^+e^-}
\newcommand{\EEMM}{e^+e^-\to \mu^+\mu^-}
\newcommand{\MM}{\mu^+\mu^-}

\newcommand{\beq}{\begin{equation}}
\newcommand{\eeq}{\end{equation}}
\newcommand{\bitm}{\begin{itemize}}
\newcommand{\eitm}{\end{itemize}}
\newcommand{\rhad}{$R_{\rm had}$}
\newcommand{\Rhad}{R_{\rm had}}
\newcommand{\rmm}{R_{\mu \mu}}

\newcommand{\amu}{$a_{\mu}$}
\newcommand{\eehad}{$\EE\to {\rm hadrons}$}

\begin{document}
\begin{CJK*} {UTF8} {gbsn}

\title{\boldmath New approach to finding invisible states in $\EE$ annihilation \\ and application to BESIII data}

\author{Glennys R. Farrar}
 \email{gf25@nyu.edu}
 \affiliation{Center for Cosmology and Particle Physics, New York University, New York, NY, 10003, USA} 

\author{Qi-Ming Li (李启铭)}
 \email{liqiming@ihep.ac.cn}
 \affiliation{Institute of High Energy Physics, Chinese Academy of Sciences,
 Beijing 100049, China}
 \affiliation{University of Chinese Academy of Sciences, Beijing 100049, China}

 \author{Chang-Zheng Yuan (苑长征)}
 \email{yuancz@ihep.ac.cn}
 \affiliation{Institute of High Energy Physics, Chinese Academy of Sciences,
 Beijing 100049, China}
 \affiliation{University of Chinese Academy of Sciences, Beijing 100049, China}

\begin{abstract}

We compare precision $\EE\to \MM$ cross section measurements by BESIII in the $E_{\rm cm}=3.8-4.6$~GeV range, to predictions based on measured \rhad\ data.  The consistency is poor (p-value $<0.012$).  Allowing for resonance contributions not seen in \rhad\ gives an excellent fit, with the state at 4421~MeV ($4.6\sigma$) giving insight into the $\psi(4415)$ and the $3.1\sigma$ structure at 4211~MeV, if confirmed, being a new, very narrow resonance. This analysis shows the power of precision $\EE\to \MM$ measurements to uncover or probe otherwise difficult to access states.  
\end{abstract}


\date{\today}

\maketitle
\end{CJK*}

Certain types of final states in $\EE$ collisions would not have been detected in existing $\EE$ experiments, due to event acceptance requirements~\cite{f_g-2_22}.  We call such final states ``invisible" for the purposes of this Letter.  In addition to truly invisible Beyond the Standard Model states, hadronic states in which less than $\approx$30\% of the energy is in visible hadrons or leptons, or whose energy deposit is too asymmetric would generally have been rejected~\cite{f_g-2_22}. The production of invisible final states in $\EE$ annihilation can be investigated in two ways:\\
\noindent$\bullet$ Measuring the total cross section inclusively, through inclusive Initial State Radiation (ISR); this approach is limited by the relatively poor resolution of the ISR photon energy and very challenging background.  \\
\noindent$\bullet$ Measuring $\sigma_{\EEMM}$ very accurately, since the $\mathcal{O}(\alpha^2)$ correction to the Born contribution depends on the true total $\EE$ cross section through a dispersion relation~\cite{Dong+CZY20}. 

Besides the intrinsic interest in potentially overlooked states, another motivation for a search for states not detected in \eehad\ is the discrepancy between the very precisely measured anomalous magnetic moment of the muon~\cite{2023g-2} and its value predicted using dispersion relations to obtain the hadronic vacuum polarization contribution from \eehad\ data~\cite{Aoyama+20,Alexandrou+ETMC22}.  This discrepancy may be resolved by revision of $R_{\rm had} \equiv \sigma(\EE \rightarrow {\rm hadrons})/\sigma^B(\EEMM)$ in the $\rho$ region, as suggested by the new CMD-3 measurement~\cite{CMD-3,CMD-3sept23}, where $\sigma^B(\EEMM) = \frac{4 \pi \alpha^2}{3 s}$ and $\alpha$ is the QED fine structure constant. Complementary approaches to this problem are nevertheless valuable.  Furthermore, as we show here, resonances which in principle can be seen in hadronic final states of $\EE$ collisions given sufficiently accurate measurements, can in some cases be more precisely probed indirectly, through their virtual impact on $\sigma_{\EEMM}$.   

In this paper, we compare precision data on $\EEMM$ from BESIII~\cite{BESmumu20} to predictions using $R_{\rm had}$ data from the PDG~\cite{PDG}.  We find that the measured $\EE \to \MM$ cross section shows structure in the 4.2-4.5~GeV range that is in significant conflict with predictions from $R_{\rm had}$ data.  We look for  possible sources of energy dependent systematic uncertainties in the $\EE \to \MM$ measurements, but none appear capable of explaining the data.  However allowing for contributions from resonances whose contributions to $R_{\rm had}$ have gone undetected provides an excellent fit. 

\begin{figure*}[htbp]
\centering
\vspace{0.1in}
\includegraphics[trim={0.08in 0 0 0},clip,width=0.33\textwidth]{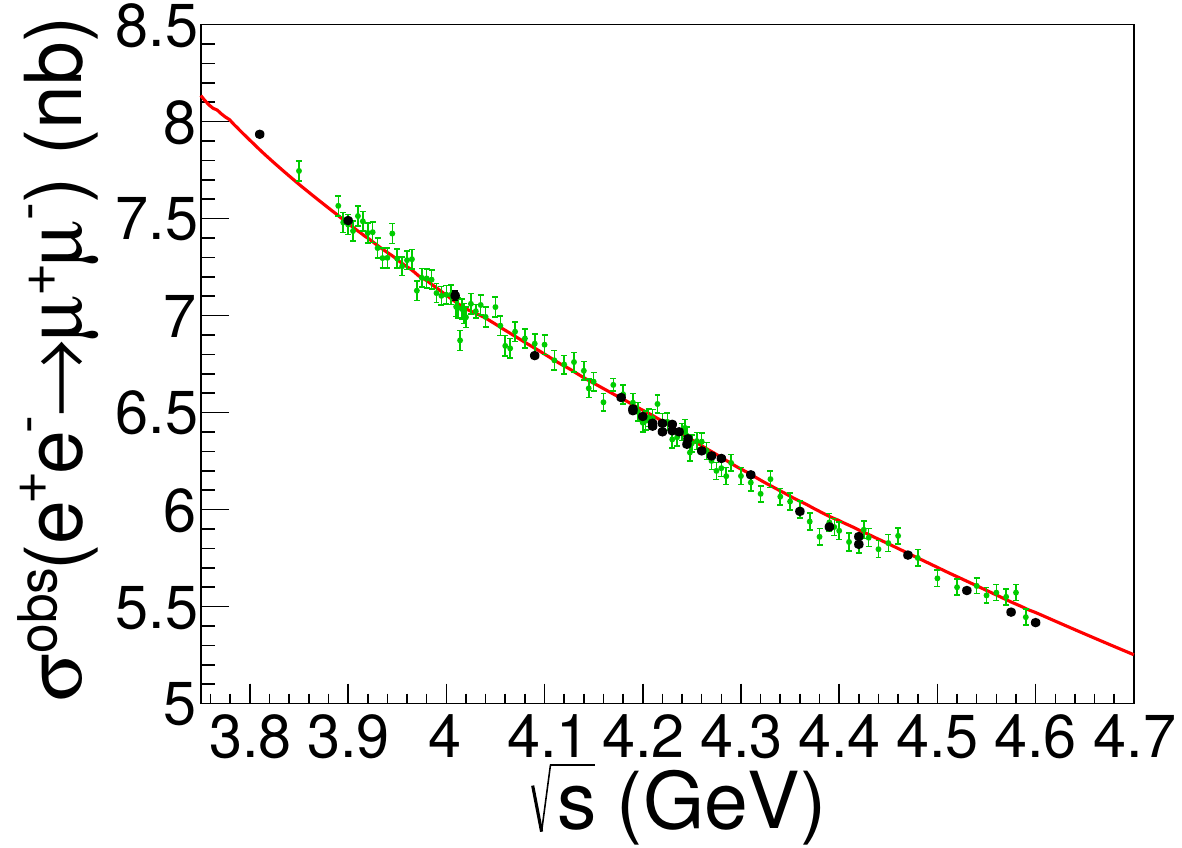}
\includegraphics[width=0.33\textwidth]{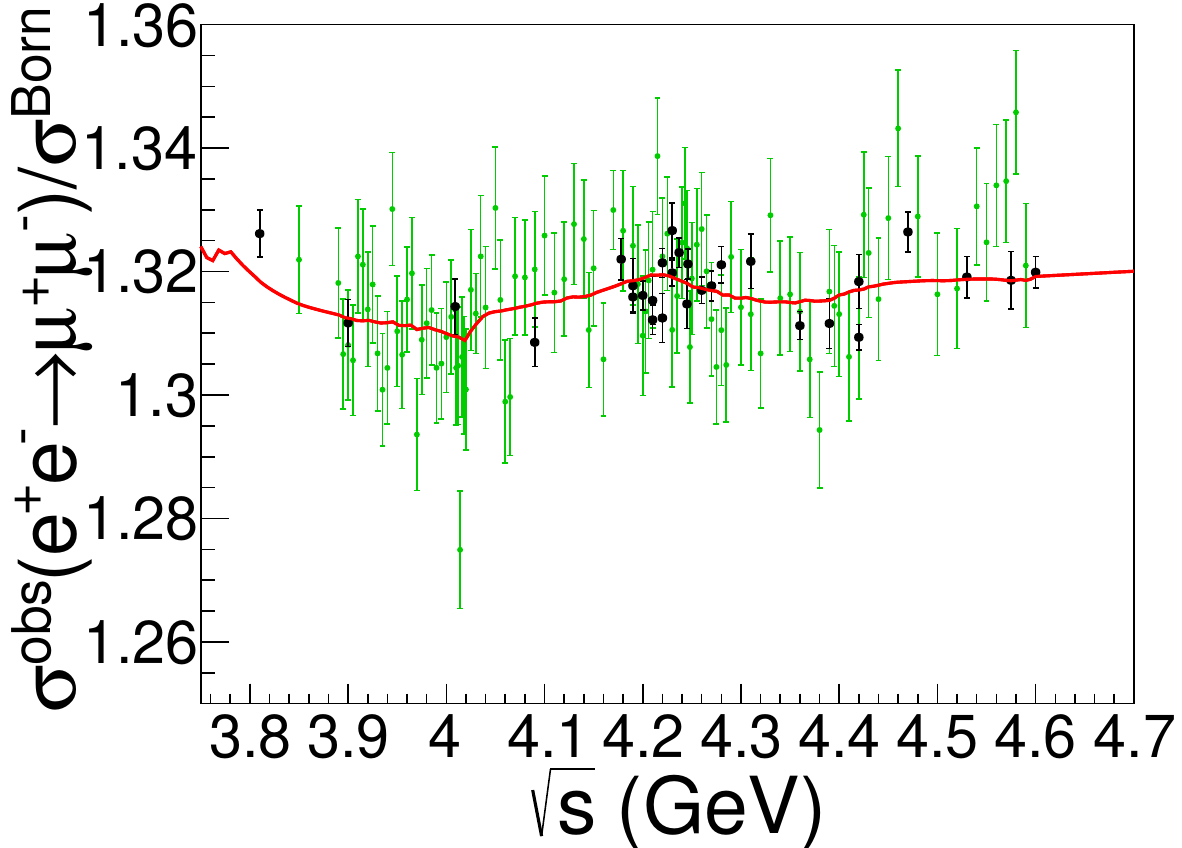}
\includegraphics[width=0.33\textwidth]{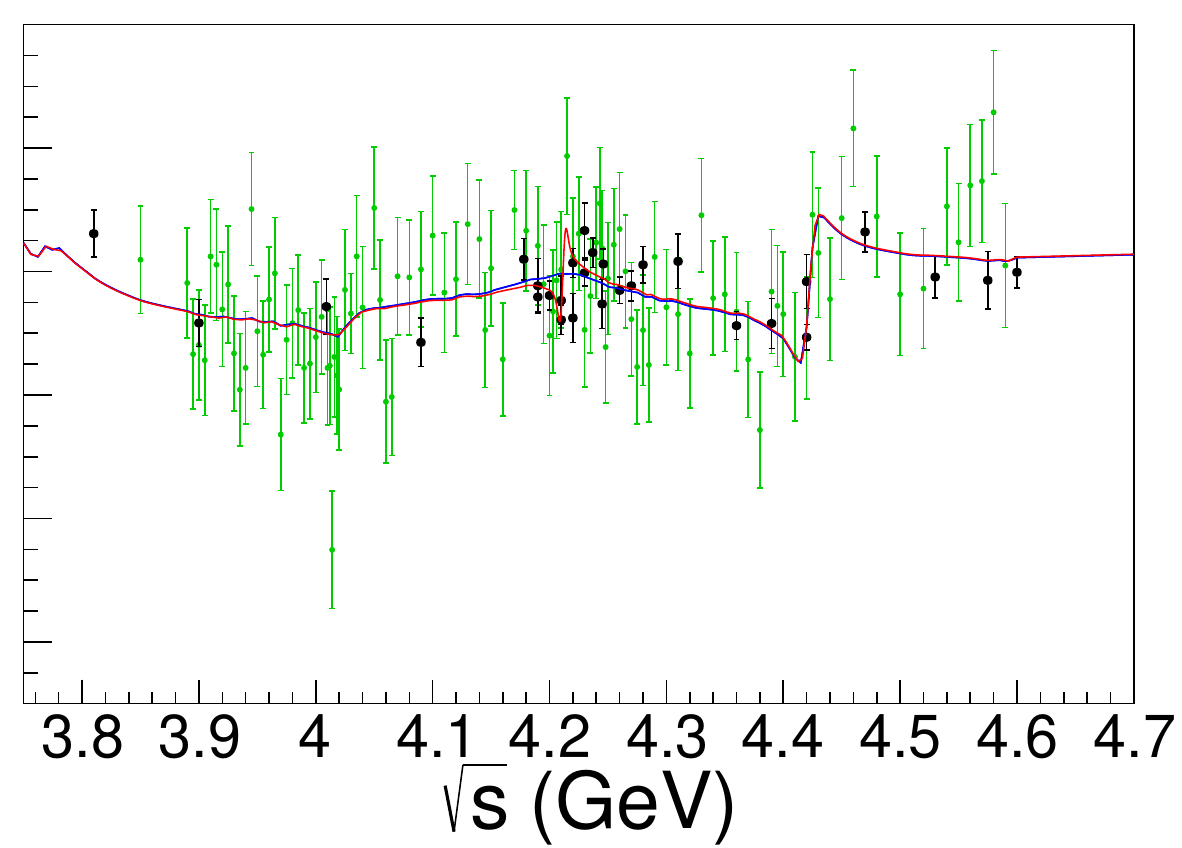}
\caption{\label{fig:compare} Left: $\sigma(\EEMM)$ measured by the BESIII experiment~\cite{BESmumu20}.  Center and Right: The black (green) points show the corresponding $\rmm$ from high-luminosity (scan) data, with statistical errors. The red curve in the center panel shows the prediction calculated from $R_{\rm had}$, shifted by the rescaling in Eq.~\eqref{eq:rescale} to account for systematics.  The right panel shows the fit with one (blue) or two (red, mostly covering blue) constant-width resonances added to the $R_{\rm had}$ data.
} 
\end{figure*}

\noindent {\bf Data used}\\ 
The BESIII collaboration measured the cross section $\sigma_{\MM}$ for $\EE\to \MM$ at center-of-mass (c.m.) energies from 3.8 to 4.6~GeV~\cite{BESmumu20}.  The raw cross section data is shown in the left panel of Fig.~\ref{fig:compare}.  Dividing the cross section by the Born cross section for easier visibility, the center panel shows $R_{\MM} \equiv \sigma(\EEMM)/\sigma^B(\EEMM)$.

Data samples were collected at 133 c.m. energies with the BESIII detector, operating at the BEPCII collider from 2011 to 2017.  The c.m. energy of each dataset is measured using dimuon events, with an uncertainty of $\pm 0.8$~MeV~\cite{bes3_ecm}.\footnote{To test for the possible impact of potential correlated and uncorrelated systematic uncertainties among data from different running years, we also perform the analysis using only data taken in winter and spring 2013/2014;  there is no significant difference. In the restricted sample, 104 energy scan data sets were taken from 3.84 to 4.59~GeV, each with an integrated luminosity of about 8~pb$^{-1}$. The scan data were taken from Dec. 9, 2013 to Jan. 24, 2014.  For each point in the scan data-taking, the accelerator energy was set to a specific value, data was taken for about 7 hours, then the energy was increased one step and data was taken again.  These data were taken with a monotonically increasing energy.  Five additional high-luminosity data sets were taken in the remainder of the same data-taking period, at c.m.~energies 4.600, 
4.467, 
4.527, 
4.575, 
and 4.416~GeV,
with integrated luminosities of 587, 49, 112, 111, and 1044~pb$^{-1}$, respectively.} 
The statistical precision of the BESIII measurements is about 0.7\% for the energy scan data sets and is from 0.16\% to 0.35\% for the other data sets. 

The systematic uncertainty of the measurement is estimated by BESIII to be about $2.91\%$ and is common to all the data sets; it originates from several sources including the luminosity measurement (1\%), uncertainty in tracking efficiency (1\%), detector acceptance, signal extraction, Monte Carlo generator, and so on.  
We allow for this with a linear rescaling of the $\EE\to \MM$ measurement, as discussed in detail below.  Our analysis concentrates on a possible shape difference between the observed and predicted $R_{\MM}$ as a function of energy, which supplies information about final states which have escaped detection in the direct measurement of the hadronic cross section. The precision of our numerical analysis is at the 0.1\% level.

\noindent {\bf Theoretical framework}\\
The total cross section $\sigma_{\MM}$, for the exclusive process $\EEMM$, is related to the vacuum polarization (VP), $\Pi(s)$, by
\begin{equation}
\label{eq:sigfromPi}
\sigma_{\MM}(s) = \frac{\sigma^{\rm B}_{\MM}(s)}{|1-\Pi(s)|^2},
\end{equation}
where $s$ is the c.m. energy squared.
The charged lepton contribution to $\Pi(s)$ can be calculated in perturbation theory and we omit it in subsequent formulae for clarity. 
The contribution to $\Pi(s)$ from virtual states other than leptons, $\Pi_{nl}(s)$, is related to the total cross section $\sigma_{nl} (s)$ to not-leptons final states in the one-photon exchange approximation through a dispersion relation~\cite{Dong+CZY20}:
 \beq
\Pi_{nl}(s)= \frac{s}{4\pi^2 \alpha} \int_{4m^2_{\pi}}^{\infty}
\frac{\sigma_{nl} (s^{\prime})}{s-s^{\prime}+i\epsilon } d s^{\prime}~.
\label{dspnrelationa}
 \eeq
Using the identity
  $(x+i\epsilon)^{-1} =  P(1/x)-i\pi \delta(x)~, $
gives
 \beq
 \label{eq:RePi}
\Pi_{nl}(s)= - \frac{s}{4\pi^2 \alpha} \, P\! \int_{4m^2_{\pi}}^{\infty}
\frac{\sigma_{nl} (s^{\prime})}{s^{\prime}-s} d s^{\prime} - i
\frac{s}{4\pi \alpha} \sigma_{nl}(s)~, 
 \eeq
where $P$ represents principal value.  Following~\cite{Dong+CZY20}, the integration is performed analytically for narrow resonances $\jpsi$, $\psip$, $\Upsilon(1S)$, $\Upsilon(2S)$, and $\Upsilon(3S)$. The high energy part of the integration assumes that $R(s)=R(s_1)$ is constant above a certain value $s_1$ (taken to be above the $\Upsilon$ resonance region), and the integral between threshold and $s_1$ is carried out numerically after separation of the principle value part. Thus 
\begin{widetext}
\begin{equation}
\label{eq:RePiEval}
\rm{Re}\, \Pi_{nl}(s)= \frac{3s}{\alpha} \sum\limits_{j}
\frac{\Gamma^j_{\EE}}{M_j}\frac{s -M_j^2}{(s -M_j^2)^2 + M_j^2 \Gamma_j^2}
+\frac{\alpha}{3\pi} R(s_1)\ln\left|\frac{s-s_1}{s_1}\right|
- \frac{s}{4\pi^2 \alpha}
 \int_{4m^2_{\pi}}^{s_1} \frac{\sigma_{\rm nr} (s^{\prime})-\sigma_{\rm nr} (s)}
 {s^{\prime}-s}\mathrm{d} s^{\prime} -\frac{s\sigma_{\rm nr} (s)}{4\pi^2 \alpha}
 \ln \left|\frac{s_1-s}{4m^2_{\pi}-s}\right| ,
\end{equation}
\end{widetext}
where $\Gamma_j$, $\Gamma^j_{\EE}$, and $M_j$ denote the total width, partial width to $\EE$, and mass of the resonance $j$, respectively. Here, $\sigma_{\rm nr} (s)$ is the $\sigma_{nl} (s)$ in \eqref{eq:RePi} with the contributions from narrow resonances subtracted.

The measured cross section for $\EEMM$ includes unobserved radiated photons, so before comparing to data we integrate the exclusive cross section $\sigma_{\MM}(s)$ calculated from Eq.~\eqref{eq:sigfromPi}, with the ISR function as discussed in~\cite{Dong+CZY20}.  

We model a possible undetected contribution to $\sigma_{nl}$, either unresolved due to gaps in the $R_{\rm had}$ measurements or decaying dominantly into undetected final states, as being due to one or two resonances, R1 and R2. The generalization to more resonances is straightforward from this case, and allowing for two unseen states proves sufficient for our purposes.  For this initial study, we assume R1 and R2 can be described by constant-width Breit-Wigner functions  
\begin{equation}
   A=\frac{\sqrt{12\pi\Gamma_{ee}\Gamma} }{s-m^2+im\Gamma}~,
\label{eq:bw_c}
\end{equation}
with $m$, $\Gamma$, and $\Gamma_{ee}$ being the mass, width, and electronic partial width of the resonance.  We allow for the possibility that the more massive resonance (R2) can have common, interfering decay channels with R1.

The resonances R1 and R2 may or may not have final states contributing to the detected \rhad, but their undetected contributions simply add to $\sigma_{nl}$ without interference since the two types of final states are inherently orthogonal.  If there is only one resonance, R1, it can be treated using Eqs.~\eqref{eq:RePi} and~\eqref{eq:RePiEval}, just like the known vector mesons.  This fit entails 3 new parameters --- the mass, width and electronic width of R1: $m_1$, $\Gamma_1$, $\Gamma_{ee,1}$.  However in computing $|1 - \Pi_{R\rm had} - \Pi_R|^{-2}$, where $\Pi_R$ is the contribution from the new sector, one must take care about the proper treatment of possible interference terms within the new sector.  When there are two resonances, we have two masses, widths and $\Gamma_{ee}$'s, and in addition the branching fraction $\BR$ and relative phase $\phi$ of R2 decay into R1 final states, which we take to be energy independent constants.  If some resonance has both visible and unseen channels, the values of $m$, $\Gamma$ and $\Gamma_{ee}$ should be compatible when proper account is taken of interference within each sector; for this, a more sophisticated treatment than the constant width BW used here could be warranted. 

The contribution to $\sigma_{nl}$ of the resonances R1 and R2 not already included via the measured $R_{\rm had}$ is then, in this approximation:
\begin{align}
  \sigma_{R1+R2} = & \,| A_{R1} + e^{i \phi} \sqrt{\BR}\, A_{R2}|^2 + (1-\BR)|A_{R2}|^2 \notag \\
  \equiv & \, \sigma_{R1} + \sigma_{R2} + \sigma_{R1R2} ~, 
\label{eq:sigR1R2}
\end{align}
where $A_{R1,R2}$ are the appropriate BW functions, given in Eq.~\eqref{eq:bw_c} for the constant width treatment, and $\sigma_{R1R2}$ is merely the shorthand  
\beq
\label{eq:defsigR1R2}
\sigma_{R1R2} \equiv  2 \sqrt{\BR}\,{\rm Re} [A^*_{R1} A_{R2}\, e^{i \phi}]~.
 \eeq
The first two terms in Eq.~\eqref{eq:sigR1R2} are just Breit-Wigner cross sections and their contributions to $\Pi_{nl}(s)$ in Eq.~\eqref{eq:RePi} are treated in the standard way.  The last term in Eq.~\eqref{eq:sigR1R2}'s contribution to $\rm{Re}\, \Pi_{nl}$ in \eqref{eq:RePi} must be directly integrated.  
Taking the BW's for R1 and R2 to be momentum-independent as in Eq.~\eqref{eq:bw_c} we find
\begin{align}
\label{eq:PiR1R2}
 &{\rm Re}\, \Pi_{R1R2}(s) \equiv  - \frac{s}{4\pi^2 \alpha}\, P \! \int_{4m^2_{\pi}}^{\infty}
\frac{2 \sqrt{\BR}\,{\rm Re} [A^*_{R1} A_{R2}\, e^{i \phi}](s^{\prime})}{(s^{\prime}-s)} d s^{\prime} \notag \\
= & 
\, K \,{\rm Re} \left[ \frac{e^{i \phi}}{(s-a)(s-b)(b-a)}  \left(b \,\ln \frac{a}{s} - a\, \ln \frac{b}{s} + s \,\ln \frac{b}{a} \right)  \right],
\end{align}
where $a = m_1(m_1 + i \Gamma_1)$, $b= m_2(m_2 - i \Gamma_2)$, $K = 6 s/ (\pi \alpha) \sqrt{\BR \,\Gamma_{ee,1} \,\Gamma_1 \,\Gamma_{ee,2} \,\Gamma_2}$ and ln($|z|e^{i \theta}) = \mathrm{ln}|z| + i \,\theta $.  If the resonance contributions are not constant-width Breit-Wigners, the integral can be done numerically.

\noindent{\bf\boldmath Fitting $\EEMM$ from \rhad}\\  
We define $\chi^2$ to assess the quality of a fit as 
\begin{multline}
\label{eq:rescale}
\chi^2 = \sum_i^N  \left( \frac{R^i_{\MM}-\hat{R}^i_{\MM}/(f_{0}+f_{1}(E_i-E_{c}))}{\sigma_{R^i_{\MM}}}\right)^2  \\ +~~\frac{\frac{1}{N} \sum_i^N (1-f_{0} - f_{1}(E_i-E_{c}))^2 }{(0.0291)^2} ,    
\end{multline}
where $R^i_{\MM}$ and $ \hat{R}^i_{\MM}$ are the predicted and measured values of $R^i_{\MM}$ at the $i$th data point, $\sigma_{R^i_{\MM}}$ is the statistical error on the $i$th data point, 0.0291 is the systematic overall normalization error estimate from BESIII for these measurements, and $f_0$, $f_1$ provide a linear, energy-dependent rescaling to the overall normalization; $E_c = 4.2$~GeV is the mean energy of the dataset.  

The best fit to the $R_{\MM}$ data, using Eq.~\eqref{eq:sigfromPi} and the measured $R_{\rm had}$ given in the PDG tabulation~\cite{PDG} to evaluate $\Pi_{nl}$ from Eq.~\eqref{eq:RePi}\textbf{}, yields the fit shown in the left and center panels of Fig.~\ref{fig:compare}, with normalization rescaling parameters $f_0=1.00286\pm 0.00052$ and $f_1 = 0.01811 \pm 0.00301$ GeV$^{-1}$.   The normalization rescaling ranges from $-$0.004 to $+$0.01 over the 3.8-4.6~GeV range of the data --- well within the estimated absolute normalization uncertainty of 0.0291.  

However even after this rescaling, the energy dependent structure of the data is significantly different from the prediction based on the measured $\sigma_{e^+e^-\rightarrow {\rm had}}$, as can be seen by eye and as reflected in the goodness-of-fit parameter $\chi^2/ndf=170.47/(133-2)$, corresponding to a confidence level of only $0.012$.  Thus in the following we consider the possibility that some contributions to $\sigma_{e^+e^-\rightarrow {\rm had}}$ could have been missed.

The right panel of Fig.~\ref{fig:compare} shows the fit with one and two additional resonances.  
The fit parameters are listed in Table~\ref{tab:c}.  The fit with one additional resonance improves the $\chi^2$ by 27.6 with 3 more free parameters giving $\chi^2/ndf=142.87/(133-5)$. The statistical significance of this resonance is $4.6\sigma$. However the confidence level is $0.17$, indicating possible room for improvement in the fit to the data.  

\begin{table}[htbp]
\caption{Fit results adding contribution from constant-width Breit-Wigner resonances;
``---'' means ``not applicable" and ``fixed" means these parameters were fixed at the one BW value.}
    \label{tab:c}
    \centering
    \begin{tabular}{l|c|c}
    \hline\hline
      Parameters  & one-BW fit &  two-BW fit \\ \hline
      $m_1$ (MeV)           & ---      & $4211.4\pm 2.6$   \\
      $\Gamma_1$ (MeV)      & ---    & $  0.14\pm 0.16$   \\
     $\Gamma_{ee,1}$ (keV) & ---  & $ 0.076\pm 0.045$  \\
      $m_2$ (MeV)           &  $4421.5\pm 3.4$& 4421.5 (fixed)  \\
      $\Gamma_2$ (MeV)      &  $15.9\pm14.7$ &  15.9  (fixed) \\
     $\Gamma_{ee,2}$ (keV) &  $0.63\pm0.31$    & 0.63  (fixed) \\
      $f_{0}$                   & $1.0019\pm 0.0006$ & $1.0019$ (fixed) \\
      $f_{1}$ (GeV$^{-1}$)                   & $0.0175\pm 0.0032$ & $0.0175$ (fixed) \\\hline
      $\chi^2/ndf$          & $142.87/(133-5)$    & $127.74/(133-8)$    \\ \hline\hline
    \end{tabular}
\end{table}

\begin{figure*}[htbp]
\includegraphics[width=0.45\linewidth,height=4.5cm]{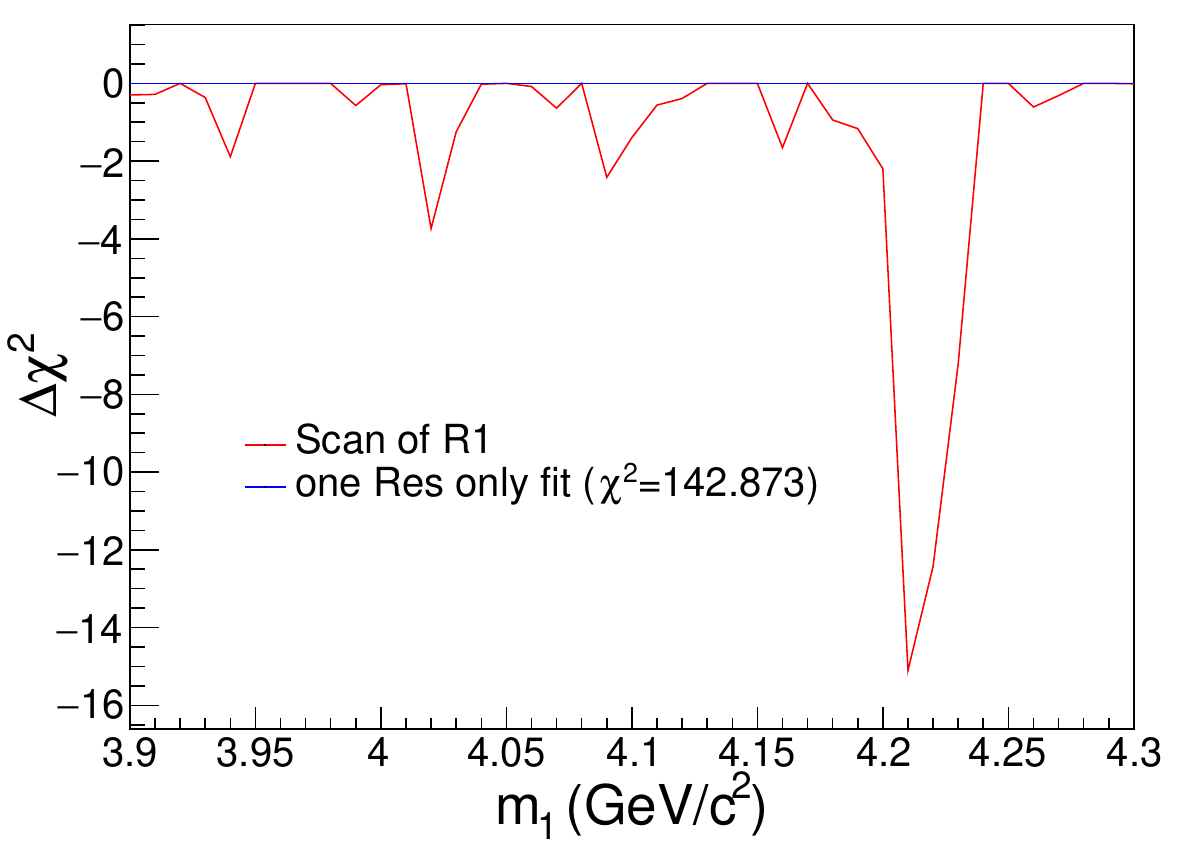}
\includegraphics[width=0.45\linewidth,height=4.5cm]{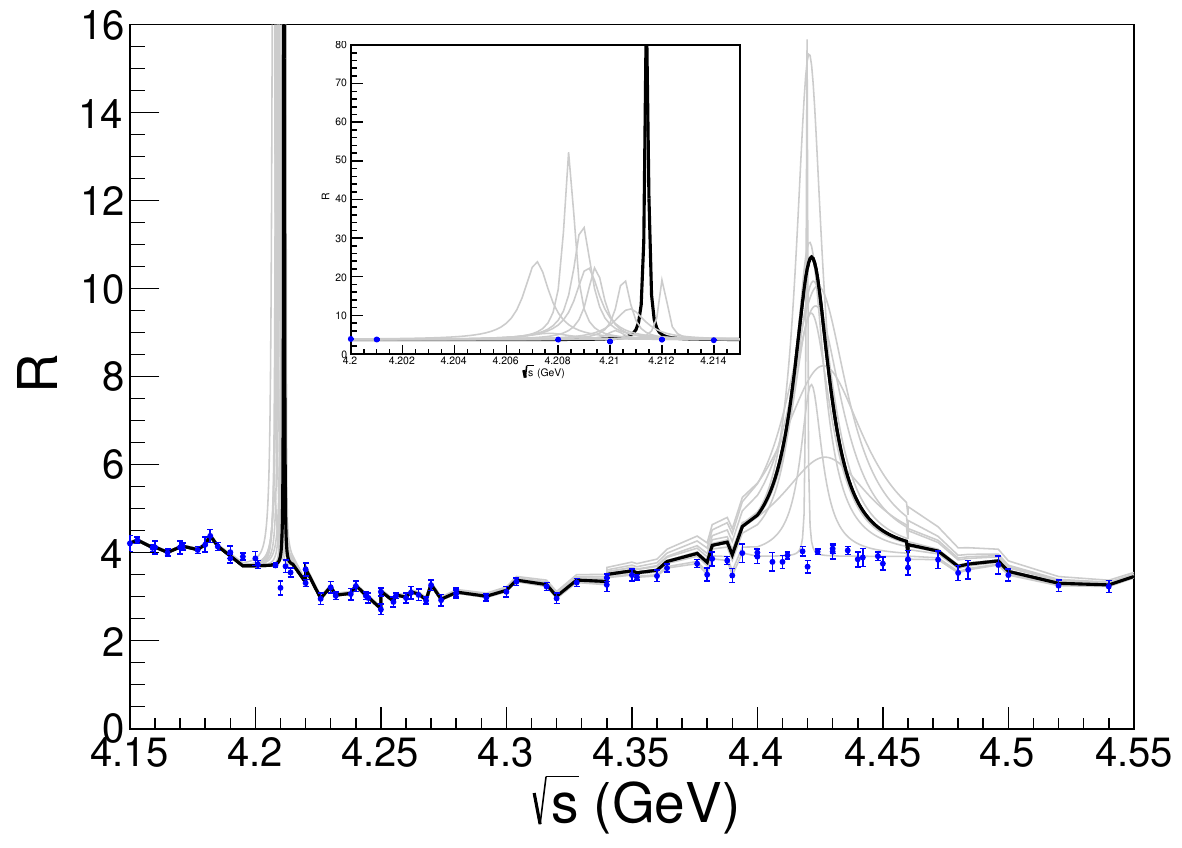}
\caption{Left: Variation of $\chi^2$ when scanning over the mass of a second contributing resonance. Right: PDG data on $R_{\rm had}$ (blue), with the contribution of R1 and R2 added shown in black (connected with a line for readability). The grey curves are 10 sample fits. \label{fig:R} 
} 
\end{figure*}

The fit with two additional resonances has a $\chi^2/ndf=127.74/(133-8)$, corresponding to a confidence level of $0.42$, indicating a very good fit to the data.  The masses of the two resonances are too far apart relative to their narrow widths to interfere, so we drop $\mathcal{B}$ and $\phi$ for the final fitting.  Allowing for a second resonance improves the $\chi^2$ by 16.1 with 3 more free parameters compared to the fit with one resonance, as can be seen in the left panel of Fig.~\ref{fig:R} which shows the variation in $\chi^2$ when scanning over the mass $m_1$ of a possible additional resonance taking the parameters of R2 to be fixed.  The statistical significance of the second resonance is $3.1\sigma$. Figure~\ref{fig:R}(right) shows the contribution of R1 and R2 to the $R$ value, where $R$ here means the total $\EE$ cross section to states other than charged lepton pairs, divided by $\sigma^{\rm B}_{\MM}$. Inclusion of a third resonance does not improve the fit significantly, as seen in the left panel of Fig.~\ref{fig:R}; the most significant (but only $0.8\sigma$) contribution is at around $4020$~MeV.

As an alternative to the procedure above, in which we used an energy-dependent rescaling to account for systematic errors, one could require the systematic rescaling to be an energy-independent constant shift.  Without invisible resonance contributions, the best fit to the $R_{\MM}$ data gives in this case $f=1.00369$ with $\chi^2/ndf$=242.95/(133-1) and CL=$1.4\times10^{-8}$. 
Fitting with two additional resonances, the lower-mass resonance is similar to R2 in the baseline fit, while the higher mass one is at 5.6~GeV and has a width of 1.7~GeV.  The upper resonance improves the $\chi^2$ by 59 with 3 more parameters, but 
given its unphysically large width and mass far above the energy of the last data point, we conclude that functionally this resonance is providing the energy-dependent rescaling of the data called for in the baseline fit and we reject the resonance interpretation on physical grounds.  


\noindent{\bf Implications for the vacuum polarization}\\
An important element of the Standard Model prediction of the muon anomalous magnetic moment is the determination of the Hadronic Vacuum Polarization (HVP) contribution, from experimental measurements of \eehad\  in the ``R-ratio" method. 
The HVP contribution to the muon g-2 is 
\begin{equation}
\label{eq:amu}
 a_\mu^{\rm HVP, LO} = \frac{\alpha^2 }{3 \pi^2} \int ds \frac{R(s) K(s) \Theta(s)}{s}  ~,
\end{equation}
where $s$ is the c.m. energy squared, $R(s) \equiv \sigma(e^+e^- \to  {\rm hadrons}) / \sigma_{\rm B}(\EE\to \MM)$ and the function $K(s)$ is given in~\cite{Aoyama+20}. The function $\Theta(s) \to 1$ when calculating $a_\mu^{\rm HVP, LO}$, but we include it for more detailed comparison to lattice QCD calculations below.  With a different weighting function an analogous calculation yields the running of the QED coupling to higher scales.  

Modifying the vacuum polarization to include invisible or not-yet-detected contributions to the observed R-ratio, whether from unseen hadrons or new physics (HVP being a minor misnomer in the latter case), modifies the predicted value of g-2. 
Reference~\cite{Aoyama+20} performed a comprehensive analysis of experiments on $e^+ e^- \rightarrow \mathit{hadrons}$ and reported an HVP contribution $a_\mu \equiv \rm{(g-2)_{\rm had}}\times 10^{10} = 693.1 \pm 4.0$.   This leads to a Standard Model prediction $5.0\sigma$ less than the latest muon g-2 measurement~\cite{2023g-2}.  However CMD-3~\cite{CMD-3,CMD-3sept23} recently reported a new determination of $R_{\rm had}$ in the $\rho$ resonance region to much higher accuracy than previous experiments, with improved control of systematics and using more sophisticated radiative corrections.  Adopting the CMD-3 results gives $a_\mu({\rm exp})- a_\mu({\rm pred})= 4.9 (5.5)\times 10^{-10}$~\cite{CMD-3sept23}.  

We calculate the contributions of the new resonances to \amu\ and $\Delta\alpha(Z)$, by sampling the fit parameters according to the covariance matrix.  The contributions of resonances R1 and R2 to $a_\mu^{\rm HVP}$ and $\Delta\alpha(Z)$ are closely correlated to one another and make only a small change to the predictions: an increment of $\delta a_\mu = 2.18^{+2.16}_{-0.10}\times 10^{-11}$ and $\delta \Delta\alpha(Z) =0.66^{+0.32}_{-0.19} \times 10^{-4}$.  For more direct comparison to the predictions of lattice QCD, we use window functions to calculate the R1 and R2 contributions to the intermediate and short distance window functions~\cite{colangelo+22}, finding: $\delta a_\mu^W = 0.314^{ +0.279}_{-0.165}\times 10^{-11}$ and $\delta a_\mu^{SD} = 1.86^{ +1.63}_{ -0.98}\times 10^{-11}$. 

\noindent{\bf Interpretation}\\
The dependence of $\sigma(\EE \to \MM)$ on energy seen in the high-precision measurements by the BESIII collaboration~\cite{BESmumu20}, from $E_{\rm cm} = 3.8$ to  $4.6$~GeV, is poorly described using dispersion relations and the measured $\sigma(\EE \to {\rm hadrons})$.  The shape of the spectrum has a confidence level of only $0.012$, after allowing for a linear, energy-dependent shift in absolute normalization within the systematic uncertainty estimate.  Adding additional resonances, R1 and R2, results in a very good fit.  
The contributions of R1 and R2 to $\Rhad$ required to explain the $\EEMM$ cross section are shown in the right panel of Fig.~\ref{fig:R}, with the mean value in black and 10 random realizations from the covariance matrix in grey.  The $\Rhad$ data in blue shows no such structure.  This means that if the R1 and R2 signals in $\EEMM$ are real, their final states somehow elude detection.  

The upper state R2, whose statistical significance is $4.6\sigma$, has mass $4421.5 \pm 3.4$~MeV, width $15.9\pm 14.7$~MeV and $\Gamma_{ee} = 0.63\pm 0.31$~keV. The mass is virtually identical with that of $\psi(4415)$ whose PDG value is $4421\pm 4$~MeV and the width is narrower than that of $\psi(4415)$ by $46\pm 25$~MeV, slightly less than $2\sigma$. A possible interpretation of the combined $R_{\MM}$ and $R_{\rm had}$ data is that R2 is in fact the $\psi(4415)$, with a width $\lesssim 30$~MeV to account for the structure in $R_{\MM}$, and a substantial fraction of its final states not being detected in the $\Rhad$ measurements, for some reason. In this scenario, the $\Gamma_{ee}$ of R2 should be interpreted as $\Gamma_{ee}\times \BR(\psi(4415)\to {\rm invisible})$ and that of the $\psi(4415)$ from PDG ($0.58\pm 0.07$~keV)~\cite{PDG} as $\Gamma_{ee}\times \BR(\psi(4415)\to {\rm visible})$ final states.  Then $\Gamma_{ee}$ of the $\psi(4415)$ would be about 1.2~keV. It should be noted that interpretation of the measurements is quite model dependent~\cite{moYuan10} and that the resonance parameters of the $\psi(4415)$ in PDG are estimates~\cite{PDG}. 

The lower mass resonance, R1 at $4211.4\pm 2.6$~MeV, with $\Gamma = 0.14\pm 0.16$~MeV and $\Gamma_{ee} = 0.076\pm 0.045$~keV has a statistical significance of $3.1\sigma$ in the $R_{\MM}$ data. It is very different from the known states in this energy region, the $\psi(4160)$ and $\psi(4230)$, in both mass and width~\cite{PDG}. Thus if R1 is not just a statistical or systematic artifact, it is unlikely to be the same state as either $\psi(4160)$ or $\psi(4230)$. Its width from $\EEMM$ is compatible with being so narrow it simply lies between energy scan points. If that were the case it could decay into conventional, visible final states and be detected directly by rescanning that energy range; the only puzzle would be what produces such a narrow resonance. 

\noindent{\bf Summary and Conclusions}\\
Following up on the remark that certain types of hadronic final states fail to meet the event selection criteria for \eehad~\cite{f_g-2_22}, we have performed the first search for evidence of missed states via their virtual effect in $\EEMM$. In the dispersive representation of the vacuum polarization, a narrow resonance in the hadronic cross-section produces a very characteristic dip-peak shape in the $\EEMM$ cross section. This striking structure appears in two places in the BESIII energy range.  The upper one has a statistical significance of $4.6\sigma$.  If real, its position and leptonic width suggests it is due to the known resonance $\psi(4415)$, which however has a narrower width and higher peak value than apparent in the $R_{\rm had}$ data, which could be due to a significant portion of its final states going undetected. There is also an indication of the same shape structure produced by a very narrow state at 4211~MeV, albeit with only $3.1\sigma$ statistical significance.  It cannot be excluded that these structures arise from systematic glitches in the data, but the agreement of fits using different datasets taken at different times and in both scan and hi-luminosity modes, argues for their robustness.

  There are several paths to investigate and extend these results.  Recorded but not-selected events in BESIII data can be re-examined, e.g., searching for pair-produced, undetected neutrals via interactions of their decay products in the detector or via asymmetric energy deposits which caused the event class to be rejected previously. An analysis of inclusive ISR data from BaBar, Belle, and Belle~II would in principle be complementary to this approach, but given the present energy resolution of ISR photons such a search could not detect such narrow states as are accessible with our technique.  To the extent that adequate $\EEMM$ data were available at lower energy, they could be used as here to see if undetected final states contribute to the $3.8\sigma$ discrepancy between lattice QCD and dispersive $R_{\rm had}$ determinations of the intermediate-window-function-weighted HVP~\cite{blumUKQCD23}.  

BESIII is currently taking data up to 5.0~GeV and will be able to extend the energy coverage to 5.6~GeV, with improved peak luminosity, after the upgrade in 2024; these measurements should significantly improve the sensitivity to undetected states in $\EE$ annihilation in the 3.6-6~GeV regime.  

\section*{Acknowledgments}
We have benefited from helpful discussions with A.~Bondar, L.~Dixon, M.~Karliner and N.~Weiner.
This work is supported in part by the National Key Research and Development Program of China under Contract No.~2020YFA0406300, National Natural Science Foundation of China (NSFC) under contract Nos. 11835012 and 12335004. The research of GRF has been supported by National Science Foundation Grant No.~PHY-2013199 and by the Simons Foundation. 




\end{document}